\documentclass[twocolumn]{pasj00}
\tenpoint

\SetRunningHead{ K. Asada et al. }{Helical Field of 3C 273 jet}

\title{A Helical Magnetic Field in the Jet of 3C~273}
\author{Keiichi \textsc{Asada},\altaffilmark{1,2,3}
        Makoto \textsc{Inoue},\altaffilmark{2}
        Yutaka \textsc{Uchida},\altaffilmark{3}
        Seiji \textsc{Kameno},\altaffilmark{2} \\
        Kenta \textsc{Fujisawa},\altaffilmark{2}
        Satoru \textsc{Iguchi},\altaffilmark{2} 
        Mutsumi \textsc{Mutoh}\altaffilmark{2,3}
       }
\altaffiltext{1}{Department of Astronomical Science, The Graduate University for Advanced Studies, \\ 
2--21--1 Osawa, Mitaka-shi, Tokyo 181--8588}
\email{asada@hotaka.mtk.nao.ac.jp}
\altaffiltext{2}{National Astronomical Observatory, 2--21--1 Osawa,\\Mitaka-shi, Tokyo 181--8588}
\altaffiltext{3}{Department of Physics, Science University of Tokyo, \\1--3 Kagurazaka, Shinjuku-ku, Tokyo 162--8601}
\KeyWords{galaxies: active --- galaxies: jets  ---  galaxies: quasars: individual (3C 273)  --- techniques: interferometric --- techniques: polarimetric}
\Received{2002/3/5}
\Accepted{2002/5/10}
\Published{$\langle$publication date$\rangle$}
\begin{document}
\maketitle

\begin{abstract}
Both in the Faraday Rotation Measure and the intrinsic polarization angle, new features are revealed to indicate a helical magnetic field operating along the jet of the bright active galactic nuclei 3C~273. The helical field has been suggested to be related to the formation and collimation of jets by magnetohydrodynamic models.  The distribution of the RM shows a systematic gradient with respect to the jet axis, which is expected by a helical magnetic field.  In addition, the helical field can consistently explain two types in the direction of the projected magnetic field: parallel and perpendicular to the jet axis. Further, if the helical magnetic field is generated by winding up of an initial field by rotation of the accretion disk, we can uniquely determine the direction of the disk rotation, since the jet is approaching us.

\end{abstract}

\section{Introduction}

Many jets emanating from active galactic nuclei (AGN) show narrow and well-collimated structures from less than a parsec (pc) up to Mpc scales (e.g., Zensus 1997,\,references therein).  From a theoretical point of view, several models were proposed concerning the formation and collimation process of a jet.  For the formation of jets, radiative pressure models (e.g., Icke 1980, 1989; Sikora et al. 1996) and magnetic field models (e.g., Benford 1978; Blandford, Payne 1982; Uchida, Shibata 1985, 1986; Meier et al. 2001; Koide et al. 2002) have been mainly discussed.  In the latter models, the toroidal component is crucial for the collimation process, and because of the toroidal component both the formation and collimation of jets would be simultaneously explained.  High-resolution observations using Very Long Baseline Interferometry (VLBI) have recently been providing some evidence in support for the MHD models (Junor et al. 1999; Gabuzda 2000; Gabuzda et al. 2000).  It is, however, not simple to see the toroidal component or helical structure of the magnetic field by observing the projected direction of the magnetic fields.

The well-known quasar 3C 273, at a redshift of $z = 0.158$, is one of the brightest quasars. For a Hubble constant of $H_{0}$ = 100 \ km \ s$^{-1}$ \ Mpc$^{-1}$ and a deceleration parameter of $q_{0} = 0.5$, an angular resolution of one milli-arcsecond (mas) corresponds to a linear resolution of 1.86\,pc. The jet components in 3C\,273 have displayed superluminal motions (Cohen et al. 1971), with the speeds of the apparent proper motion being 4.9 to 7.7 $c$ derived from VLBI monitoring observations (Abraham et al. 1996). A VLBI Space Observatory Programme (VSOP) observation revealed a double-helical structure in the 3C\,273 jet, which was attributed to a Kelvin--Helmholtz (K--H) instability (Lobanov, Zensus 2001).

VLBI polarimetric observations have shown that the jet of 3C 273 is highly polarized (Roberts et al. 1990), which makes it suitable to study the Rotation Measure ($RM$). Observations with the Very Large Array (VLA) showed $RM$ variations of 5 rad m$^{-2}$ on the arcsecond scale (Roberts et al. 1990).  On the mas scale, however, Very Long Baseline Array (VLBA) observations revealed $RM$s of up to 2000 rad m$^{-2}$ towards the nuclear regions, while towards the jet the $RM$s are around 200 rad m$^{-2}$ (Taylor 1998). The high $RM$s towards the core are probably caused by the magnetic field and plasma in the Narrow Line Region (NLR) around the core (Taylor 1998). 

To investigate the 3D structure of the magnetic field, a detailed analysis of both the $RM$ and the polarization angle ($PA$) should be useful. The $RM$ is related to the electron density, $n_{\rm e}$, and the magnetic field component parallel to the line of sight, $B_{||}$, as $RM \sim \int_{\rm LOS} n_{\rm e} B_{||} dr$, where $\int_{\rm LOS} dr$ represents integration along the line of sight. In an optically thin plasma, the projected direction of the magnetic field (a component perpendicular to the line of sight) is perpendicular to $PA$. 

Since $RM$ is measured by the polarization angles ($PA_{i}$) at different wavelengths ($\lambda_{i}$) through the equation $PA_{i} = PA_{0} + RM \lambda_{i}^{2}$, a more accurate $RM$ can be measured at longer wavelengths. When the $RM$ is very large, however, longer-wavelength observations may result in aliasing due to the 180$^{\circ}$ ambiguity of $PA$. To avoid this, a combination of observations at either (i) both long and short wavelengths or (ii) multiple longer wavelengths separated by only a small wavelength difference is required. Therefore, a proper selection of observing the frequencies is essential in deriving the $RM$ distribution of jets. VLBA polarimetry could thus give one of essential information as to whether or not the magnetic field plays a dominant role in jet formation and/or collimation.

\section{Observations and Data Reduction}

We analyzed polarimetric observations of the 3C\,273 jet based on archival VLBA data. Observations were carried out on 1995 December 9 at 4.702, 4.760, 4.890, and 4.990 GHz in the 5\,GHz band and 1995 November 22 at 8.102, 8.240, 8.420, and 8.590 GHz in the 8\,GHz band using all ten stations of the VLBA. Each IF has an 8-MHz bandwidth. The target of these observations was M\,87, with 3C\,273 being observed as a calibration source. Both left and right circular polarizations were recorded at each station. An {\it a priori} amplitude calibration for each station was derived from a measurement of the antenna gain and system temperatures during each run. Fringe fitting was performed on each IF and polarization independently using the AIPS task FRING.  After deriving the delay and rate difference between parallel-hand cross correlations, the cross-hand correlations were fringe fitted to determine the cross-hand delay difference.  Once the cross-hand delay difference was determined, the instrumental polarizations of the antennas were determined for each IF in the same bands with an unpolarized source, OQ 208, using the AIPS task PCAL. The polarization angle offset at each station was calibrated using data of OJ 287 observed by the University of Michigan Radio Astronomy Observatory (UMRAO). We note that the difference in the observing dates by UMRAO and VLBA is within 1 day and 5 days for 5 and 8 GHz, respectively; thus, the chance of variability between the two observations is small. Because the total polarized flux of OJ\,287 measured by the VLBA is 31.8 $\pm$ 1.7 mJy at 5 GHz and 26.4 $\pm$ 1.3 mJy at 8 GHz, while the polarized fluxes measured by the UMRAO are 32.7 $\pm$ 6.1 mJy and 24.9 $\pm$ 13.7 mJy, respectively, the corresponding values are equal within the errors, confirming that a very high fraction of the integrated polarized flux was present on VLBA scales.  We applied the same position angle calibrations for each IF within the 5 GHz and 8 GHz bands. The integrated $RM$ of OJ\,287 based on VLA observations is small, +31 rad m$^{-2}$ (Rudnick, Jones 1983). The corresponding differences in the $RM$ for different frequencies within the 5 GHz and 8 GHz bands are negligible, justifying our use of the same $PA$ calibration for all IFs at each frequency. We also note that even if a small offset in the $RM$ were introduced, it would be constant all over the source, and could not give rise to structure in the $RM$ distribution.

Images were initially obtained using DIFMAP, then imported into AIPS to self-calibrate the full datasets using the task CALIB before the final DIFMAP image.  In order to obtain the distributions of $RM$ and $PA$, we tapered images at higher frequencies to match the resolution of the lowest frequency observation. The restored beam size was 3.4 mas $\times$ 1.4 mas with the major axis at a position angle of $-$3.$^{\circ}$7. In order to register images at different frequencies, we refer to optically thin components of jet. As a result, the error of alignment was estimated to be within 10

\section{Results}

\subsection{Total Intensity and Linear Polarization Images}

We show images of the polarized intensities overlaid on the total intensity images at 5 and 8 GHz in figures 1a and b, respectively. The core is located at the northeast end of the jet, and the jet extends along a position angle of 234$^{\circ}$. The counter-jet is invisible, presumably due to Doppler de-boosting. The core is weakly polarized at the level of $\sim 0.3 \%$ and $\sim 0.8 \%$ at 5 and 8 GHz, as seen in previous observations (Lepp\"anen et al. 1995; Roberts et al. 1990; Taylor 1998). The fractional polarization of component C3 at 8 GHz is 8\%, while that at 5 GHz is 0.5\%. This suggests that this component is depolarized; however, there are no obvious signs of depolarization within the 5 GHz and 8 GHz bands. Component C3 has a flat spectrum with $\alpha = -0.13$ ($S_{\nu}\sim \nu^{-\alpha}$) between 5 GHz and 8 GHz.

\subsection{Projected Magnetic Field and Rotation Measure}

We show the distribution of the projected magnetic field, which is obtained by de-rotating the $PA_{i}$ to remove the effect of the $RM$, superposed on the total intensity contours in figure~2.  The projected magnetic field initially runs roughly parallel to the jet, then turns oblique to it at the southwestern edge of C2.

The $RM$ distribution is shown in pseudo color in figure~3. Regions of lower fractional polarization could not give a reliable $RM$ distribution because of the short observation time. The $RM$ values on either side of the jet differ significantly from each other. This gradient clearly appeared in independent observations at 8 GHz and 15 GHz (Zavala, Taylor 2001). In figure~4, we show the $RM$ distribution along the cross-section slice AB at a position angle of 36$^{\circ}$. The $RM$s differ by up to 210 rad m$^{-2}$, where the error is less than 40 rad m$^{-2}$ and is typically 10 rad m$^{-2}$.

\section{Discussion}

\subsection{Implication of the Distribution of RM}

The variation can be confidently associated with the jet itself, since it is unlikely that a foreground Faraday screen would produce such a strong fine structure comparable to the jet width (Taylor 1998). It is possible that the distribution of $RM$ could be due to a gradient in the gas density across the jet. However, in this case, we would expect to see some evidence of an interaction between the jet and this denser gas, such as the development of a sheath of longitudinal magnetic field due to shear (cf. Attridge et al. 1999) or sharp bending of the jet (cf. Nan et al. 1999). Another interesting possibility is that the variation in the $RM$ is related to differences in the component of the magnetic field along the line of sight.

Indeed, the systematic distribution of $RM$ across the jet can be very naturally interpreted in terms of a helical magnetic field. Let us consider the simple case of the side view of a straight jet (a viewing angle is 90$^{\circ}$) with a helical magnetic field along the jet. The sign of $B_{||}$ differs on the two sides of the jet as the magnetic field reverses direction. We would see an anti-symmetric distribution of the $RM$ across the jet. In this simple case, the $RM$ would be zero at the center of the jet, because there is no $B_{||}$ component along the line of sight. When the viewing angle decreases, an anti-symmetric distribution would remain, with only the addition of an offset in the absolute value of the $RM$. In fact, the viewing angle to the jet of 3C\,273 must be small, as jet components show superluminal motion, with a viewing angle of $<$ 16$^{\circ}$, estimated from a kinematical analysis (Abraham et al. 1996). The distribution is anti-symmetric across the jet with the $RM$ biased by 330 rad m$^{-2}$ (see figure~4). If the viewing angle is smaller than the pitch angle, it is possible that $B_{||}$ always has the same sign. To illustrate this, we show the line-of-sight component of the helical magnetic field across the jet at several viewing angles in figure~5. The observed bias is ascribed to the longitudinal component of the helical magnetic field with a small viewing angle. If this is the case, the toroidal component of the helical field is twisted like a right-hand screw as the jet moves downstream. We note that the helical magnetic field can naturally produce the double-helical structure revealed by the VSOP observation (Lobanov, Zensus 2001), because the helical magnetic field thread the emitting plasma and are tied to it (frozen-in). VSOP monitoring observations would be able to discriminate between these two possibilities: a helical field or a K--H instability.

\subsection{Implication of the Projected Magnetic Field}

The observed change in the direction of the projected magnetic field along the jet, as shown in figure~2, can be explained by a change in the pitch angle of the helical field and incoherent polarization radiation. Again, let us consider the simple case when we look at a helical magnetic field in an optically thin jet from the side view. The vector addition of the magnetic field integrated across the jet is always directed along the jet, independent of the pitch angle of the helical field. However, the polarization angle has an ambiguity of 180$^{\circ}$, because the polarization planes for two magnetic fields pointing in exactly opposite directions would be the same. Therefore, the resultant direction of the polarization angle integrated across the jet would depend on the pitch angle of the field. When the pitch angle is small, the magnetic field runs almost parallel to the jet, and the $PA$s on both the near and far sides of the jet are almost perpendicular to the jet, so that the $PA$ integrated across the jet is also perpendicular to the jet. On the other hand, when the pitch angle is large, the magnetic field at any given point is almost perpendicular to the jet, and the corresponding polarization angles are almost parallel to the jet.  Since the polarization radiation is incoherent, the vector accumulation of the two polarization radiations is almost parallel, resulting in a sudden jump in the integrated polarization angle of 90$^{\circ}$ as the pitch angle changes.  

The spectrum in the region where the projected magnetic field is perpendicular to the jet ($\perp$ region) is flatter than that in the region where the magnetic field is parallel to the jet ($\parallel$ region).  The fractional polarization is up to 25$\%$ in the $\perp$ region, compared to $\sim$10$\%$ in the $\parallel$ region. These facts, combined with the presence of the bright peak at the eastern edge of the $\perp$ region, suggest that this region is compressed by a shock.  In fact, in the next bright region, further southwest (i.e.\ the C1), the magnetic field is again parallel to the jet, followed by a further knot where the magnetic field is essentially parallel along the jet. Thus, the southwestern edge of C2 probably has a large pitch angle due to compression by the shock and, consequently, the integrated magnetic field appears to be perpendicular to the jet. In fact, the magnetic field structure is essentially the same as that in the $\parallel$ region in terms of the systematic gradient of $RM$, which again indicates the helical field structure. On the line 3 mas southwest of the line AB, which is more than one beam apart, the amount of the $RM$ gradient is 190 rad m$^{-2}$, where the error is less than 27 rad m$^{-2}$ and is typically 10 rad m$^{-2}$. This interpretation may also be applicable to the dichotomous nature of the apparent magnetic field properties in jets between quasars and BL Lacs (Gabuzda et al. 1989, 1992), which could be due simply to different pitch angles for their helical magnetic fields (cf. Gabuzda et al. 2000).

\subsection{Direction of Rotation of the Accretion Disk}

We believe that the most likely mechanism for the generation of the helical jet magnetic field is that the original magnetic field has been wound up by rotation of the accretion disk (e.g. Uchida, Shibata 1985). In this case, we can identify that the accretion disk is rotating clockwise (as we see it), because we know that the jet is approaching. Taking into account the generation mechanism of the helical magnetic field by MHD model, one would then observe a rotation in which the northern half of the disk is going away, and the southern half is approaching us. Next to M 87 (Ford et al. 1994; Harms et al. 1994; Macchetto et al. 1998) and the maser observations of a megamaser source, NGC 4258 (Miyoshi et al. 1995), we can predict the direction of rotation in the disk around a possible super-massive black hole. This can be applied to anything that is responsible for winding up the magnetic field, like a possible spinning super-massive black hole (Koide et al. 2002). It would be interesting to confirm the $RM$ distribution in a case in which the counter jet could also be observed. The distribution should be anti-symmetric with respect to the AGN core, if the helical field is generated by the rotation of the accretion disk or the spinning black hole.

\section{Conclusion}

In order to estimate the 3D structure of the magnetic field in the jet from AGN, we have derived distributions of $RM$ and projected magnetic field of 3C\,273 jet using archival data from a VLBA polarimetry observation. The systematic gradient across the jet in the distribution of the $RM$ is interpreted as the helical magnetic field that is suggested by MHD models. We derived the direction of the twist of the helical magnetic field as a right-hand screw from the sign of $RM$. Furthermore, two patterns of parallel and perpendicular to the jet axis in the projected magnetic fields could be explained well by the difference of the pitch angle of the helical magnetic field with incoherent polarization angles. This interpretation can be applied to the difference of the tendency in the direction of the magnetic field between quasars and BL Lacs. It can be naturally predicted that the direction of the rotation of the accretion disk is clockwise, if the observed helical magnetic field is formed by MHD models. All these points are in favor of a helical magnetic field operating predominantly in jet formation.

\bigskip

The authors thank D.C. Gabuzda for a critical reading of this paper and valuable comments, and P.G. Edwards and H. Kigure for helpful comments. This research has made use of data from the National Radio Astronomy Observatory (NRAO) archive, and from the University of Michigan Radio Astronomy Observatory, which is supported by funds from the University of Michigan.  The National Radio Astronomy Observatory is a facility of the National Science Foundation operated under cooperative agreement by Associated Universities, Inc.

\onecolumn

\begin{figure}
\begin{center}
\FigureFile(171.4mm,99.6mm){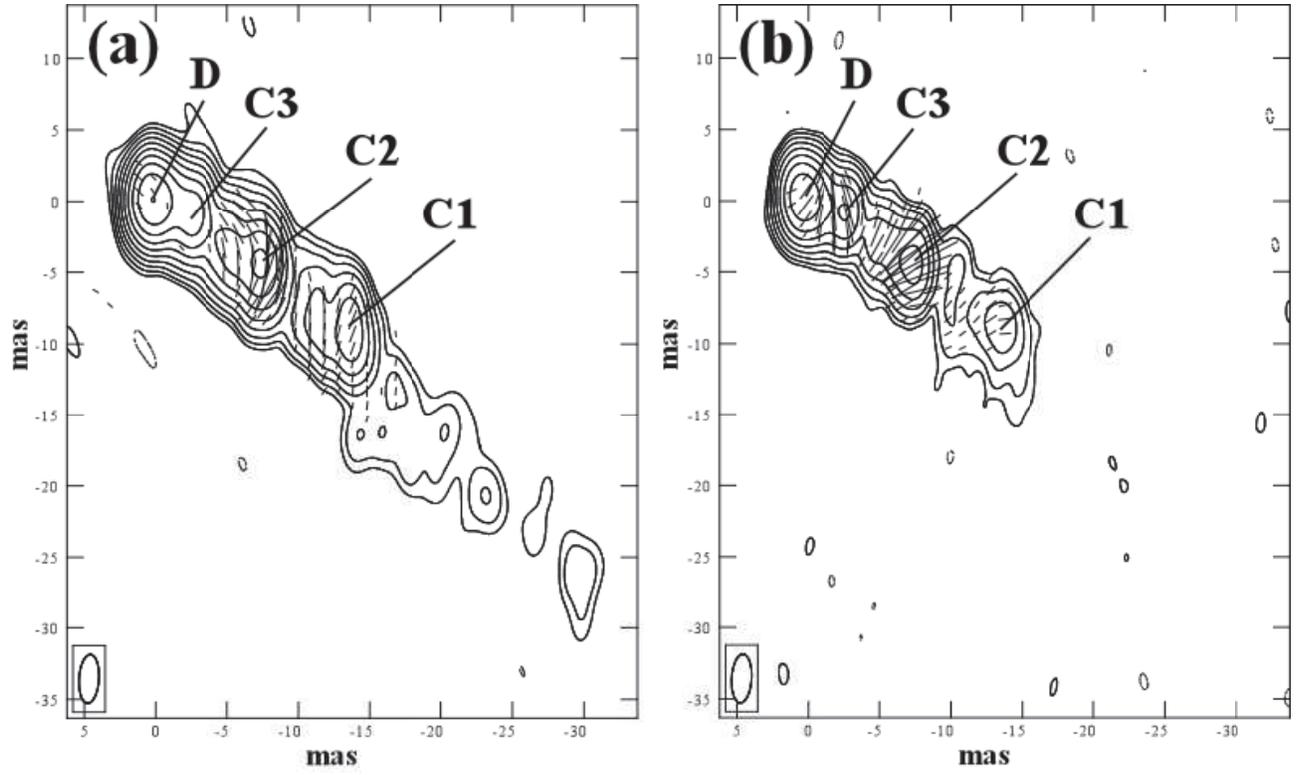}
\end{center}
\caption{Distributions of polarized intensities at (a) 4.990 GHz and (b) 8.102 GHz shown superposed on the contour images of the total intensities at each frequency. Contours are plotted at $-$1, 1, 2, 4, 8, 16, 32, 64, 128, 256, 512, and 1024 $\times$ 18.76 mJy beam$^{-1}$ and 45.36 mJy beam$^{-1}$, which are three-times the r.m.s.\ noise in the total intensities at 5 and 8 GHz, respectively.  The synthesized beam is restored by the 5-GHz beam of 3.4 mas $\times$ 1.4 mas with the major axis at a position angle of $-$ 3.$^{\circ}$7. Core (D) and jet components (C1 -- C3) are labeled by the conventional nomenclature (e.g. Lepp\"anen et al. 1995). The bars overlaid on the contours represent the directions of the polarization intensity electric vectors, the length being proportional to the polarization intensity (1 mas $=$ 50 mJy beam$^{-1}$). They are plotted in the region where the polarized intensity is greater than 3-times the r.m.s.\ noise in the polarized intensity.  The core does not show any appreciable polarization at both frequencies.}
\label{fig:PA}
\end{figure}

\begin{figure}
\begin{center}
\FigureFile(85mm,102.1mm){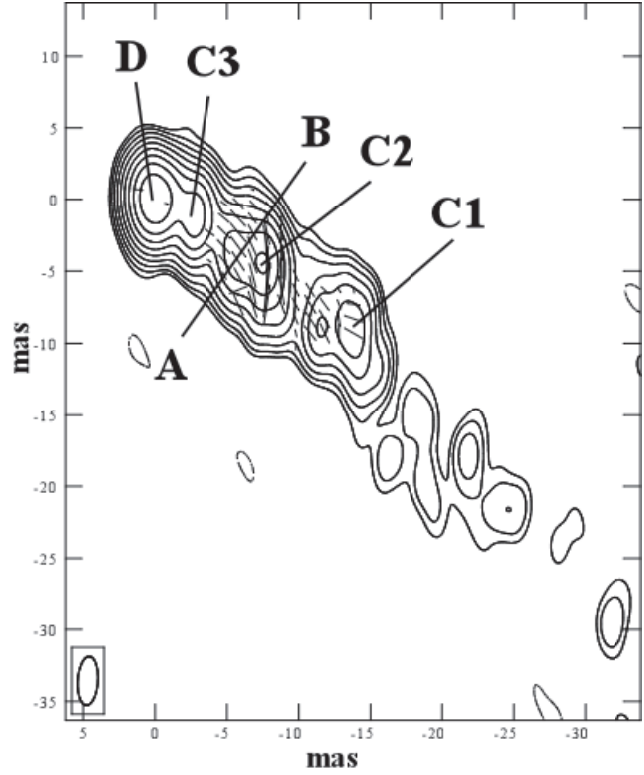}
\end{center}
\caption{Distribution of the projected magnetic field of 3C\,273 superposed on the contour image of the total intensity at 4.702 GHz. Contours are plotted at $-$1, 1, 2, 4, 8, 16, 32, 64, 128, 256, 512, and 1024 $\times$ 19.25 mJy beam$^{-1}$, which is three-times the r.m.s.\ noise in the total intensity image.  The restored beam is the same as in figure 1.  The bars overlaid on the total intensity represent the direction of the projected magnetic field derived from the intrinsic polarization angle, $PA_{0}$ (see the text), and plotted where the polarized intensity is greater than 3-times the r.m.s.\ noise in the polarized intensity at all the frequencies. In C2 a remarkable change is seen in the field direction.  The line A--B on C2 shows the position of the cross-cut shown in figure 4.  }
\label{fig:B}
\end{figure}

\begin{figure}
\begin{center}
\FigureFile(84.8mm,73.5mm){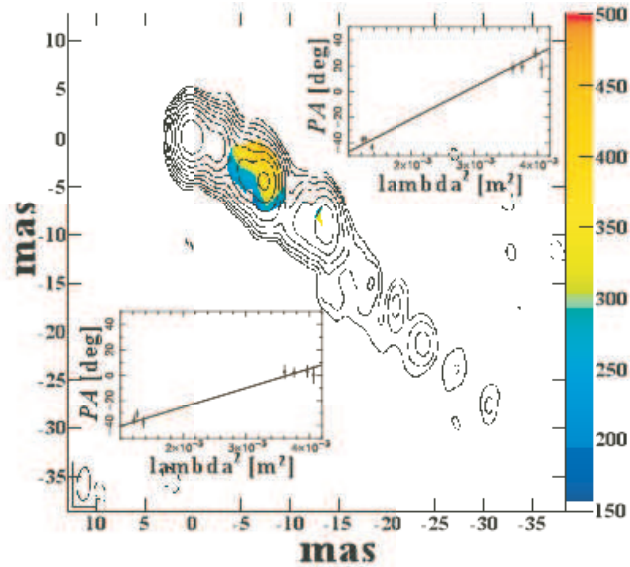}
\end{center}
\caption{Distribution of $RM$ (color scale) superposed on the total intensity image at 4.76 GHz seen in figure~2. $RM$s are plotted in regions where the polarized intensity is greater than 3-times the r.m.s. noise in the polarized intensity, but in C1 $RM$s are not shown because of a large error due to a low polarized intensity.  In C2, the $RM$ distribution is very systematic along the jet, even across the regions of different field directions.  This systematic distribution is easily understood by a helical magnetic field. We show plots of $PA$ vs. $\lambda^{2}$ for each side of the jet, where $RM$ is the maximal and minimal points on line A--B in figure~2, showing that the overall data are consistent with the behavior expected for Faraday rotation.}
\label{fig:RM}
\end{figure}

\begin{figure}
\begin{center}
\FigureFile(85mm,60.0mm){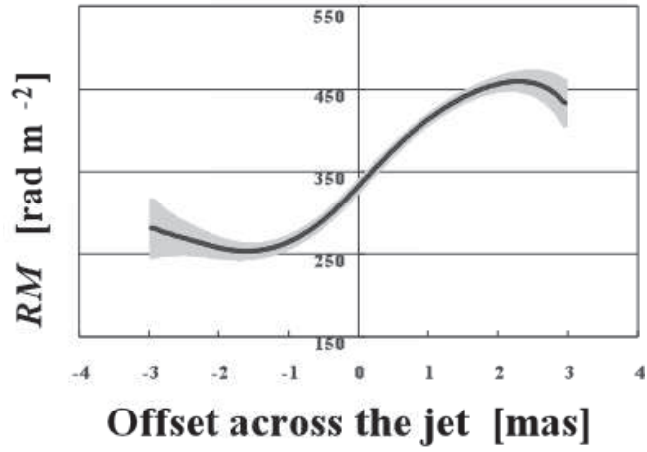}
\end{center}
\caption{Cross section of the $RM$ distribution across C2 (see line A--B in figure 2) derived using the AIPS task IMFIT.  The shaded area along the curved line of $RM$ indicates the standard deviation (1\,$\sigma$) in $RM$. The profile of the $RM$ distribution is anti-symmetric with respect to the central axis of the jet. The $RM$ variation strongly suggests the structure of the helical magnetic field.}
\label{fig:CROSS}
\end{figure}

\begin{figure}
\begin{center}
\FigureFile(85mm,58.0mm){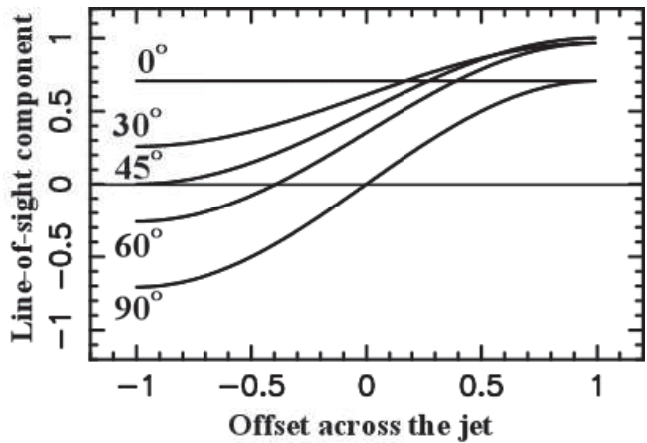}
\end{center}
\caption{Line-of-sight component of the helical magnetic field with viewing angle at 0$^{\circ}$, 30$^{\circ}$, 45$^{\circ}$, 60$^{\circ}$, and 90$^{\circ}$.  The pitch angle of the helical field is 45$^{\circ}$. The vertical scale is normalized by the strength of the magnetic field. If the viewing angle is smaller than the pitch angle, the line-of-sight component of the helical magnetic field always has the same sign.}
\label{fig:LOS_B}
\end{figure}

\end{document}